\pdfoutput=1 
\documentclass[conference]{IEEEtran}
\usepackage{amsmath,amssymb,amsfonts,chemarrow,balance}
\usepackage{amsmath} 
\usepackage{graphicx}
\usepackage{subfigure}
\usepackage{url}
\usepackage{mathenv}
\usepackage{color}
\usepackage{breqn}
 \usepackage{algpseudocode}
\newcommand{\BfPara}[1]{{\noindent\bf#1.}\xspace}
\usepackage{multirow}
\usepackage{float}
\usepackage{threeparttable}
\usepackage[inline]{enumitem}
\usepackage{wasysym}
\usepackage{etoolbox}

\usepackage[T1]{fontenc}
\usepackage[utf8]{inputenc}
\usepackage{environ}
\usepackage{tikz}
\usetikzlibrary{arrows}
\usetikzlibrary{calc,matrix}
\renewcommand{\baselinestretch}{0.975}

\usepackage{xspace}
\usepackage{setspace}
\newcommand{\note}[1]{}

\newcommand{\bc}{{Bitcoin}\xspace}

\newcommand{\cc}{{cryptocurrency}\xspace}

\newcommand*{\defeq}{\stackrel{\text{{\color{brown}def}}}{=}}
\usepackage[inline]{enumitem}

\newcommand{\etal}{{\em et al.}\xspace}

\usepackage{cite}
\usepackage{hyperref}
\def\equationautorefname~#1\null{(#1)\null}

\usepackage{mathtools}  
\usepackage{amsmath}
\usepackage{amssymb}
\usepackage{tabulary}
\usepackage{amsmath}

\usepackage[linesnumbered,ruled,noend]{algorithm2e}
\usepackage{algpseudocode}
\usepackage{amsfonts}
\usepackage{booktabs}

\colorlet{punct}{red!60!black}
\definecolor{background}{HTML}{ffffff }
\definecolor{delim}{RGB}{20,105,176}
\colorlet{numb}{magenta!60!black}

\definecolor{light-gray}{gray}{0.95}
\definecolor{darkgray}{rgb}{0.4, 0.4, 0.4}
\definecolor{editorGray}{rgb}{0.95, 0.95, 0.95}
\definecolor{editorOcher}{rgb}{1, 0.5, 0} 
\definecolor{editorGreen}{rgb}{0, 0.5, 0} 
\definecolor{orange}{rgb}{1,0.45,0.13}      
\definecolor{olive}{rgb}{0.17,0.59,0.20}
\definecolor{brown}{rgb}{0.69,0.31,0.31}
\definecolor{purple}{rgb}{0.38,0.18,0.81}
\definecolor{lightblue}{rgb}{0.1,0.57,0.7}
\definecolor{lightred}{rgb}{1,0.4,0.5}
\usepackage{upquote}
\usepackage{listings}

\begin{document}

\title{Countering Selfish Mining in Blockchains}

\author{
\IEEEauthorblockN{Muhammad Saad}
\IEEEauthorblockA{University of Central Florida\\
saad.ucf@knights.ucf.edu}
\and
\IEEEauthorblockN{Laurent Njilla}
\IEEEauthorblockA{Air Force Research Laboratory\\
laurent.njilla@us.af.mil} 
\and
\IEEEauthorblockN{Charles Kamhoua}
\IEEEauthorblockA{Army Research Laboratory\\
charles.a.kamhoua.civ@mail.mil}
\and 
\IEEEauthorblockN{Aziz Mohaisen}
\IEEEauthorblockA{University of Central Florida\\
mohaisen@ucf.edu}}

\maketitle

\begin{abstract}

Selfish mining is a well known vulnerability in blockchains exploited by miners to steal block rewards. In this paper, we explore a new form of selfish mining attack that guarantees high rewards with low cost.  We show the feasibility of this attack facilitated by  recent developments in blockchain technology opening new attack avenues. By outlining the limitations of existing countermeasures, we highlight a need for new defense strategies to counter this attack, and leverage key system parameters in blockchain applications to propose an algorithm that enforces fair mining.  We use the expected transaction confirmation height and block publishing height to detect selfish mining behavior and develop a network-wide defense mechanism to disincentivize selfish miners. Our design involves a simple modifications to transactions' data structure in order to obtain a ``truth state'' used to catch the selfish miners and prevent honest miners from losing block rewards.
\end{abstract}

\section{Introduction}\label{sec:introduction}
Blockchain technology has many applications, such as cryptocurrencies~\cite{MauriCD18,DanezisM16,SaadM18}, smart contracts~\cite{kosba2016hawk,BhargavanDFGGKK16}, Internet of things~\cite{JesusCAR18,SharmaSJP17}, health care \cite{GuoSZZ18,Rakic18}, and supply chain management~\cite{EljazzarAKE18}. Blockchain applications use a constantly evolving distributed ledger that is capable of developing consensus in a decentralized environment. As the name suggests, a blockchain is a sequence of data blocks that are cryptographically chained to one another through  one-way hash function. With the help of these mathematical constructs, blockchains employ an append-only model to prevent data tempering and preserve uniform consensus among peers in the network. 

Despite these promising capabilities, blockchains are vulnerable to a series of attacks that evolve from its design constructs, its underlying peer-to-peer network, and the applications that make use of this technology. Some of the well known attacks on blockchains include the selfish mining attacks \cite{eyal2014majority}, block withholding attacks \cite{kwon2017selfish}, the majority attack \cite{bastiaan2015preventing}, distributed denial-of-service (DDoS) attacks \cite{saad2018poster}, among others. The feasibility of each attack in blockchain applications varies depending on the network topology, adversarial requirements, peer behavior, and incentives. For example, in cryptocurrencies, it is more feasible to launch DDoS attack to exploit block size than to launch the 51\% attack \cite{BaqerHMW16}. Per Baqer \etal \cite{BaqerHMW16}, the block size can be exploited by choking the network by excessively generating low-cost transactions. On the other hand, to launch the 51\% attack, an adversary needs to acquire more than 50\% of network's hash rate to permanently gain control over the system. 

Selfish mining is one such an attack that is considered to be infeasible in practice due to centralization of the mining pools, and the potential diminishing returns. However, in this paper, we argue that selfish mining may be more viable than commonly believed, and so the existing countermeasures are insufficient to prevent selfish mining attacks. We draw attention towards recent developments in the blockchain systems that have opened new attack avenues for selfish miners. Attackers may lease sufficient hash rate from online services to combine 51\% attack with selfish mining and compromise the blockchain applications without being detected. In this paper, we also describe a threat model that partakes these new attack opportunities with a baseline attack procedure. We supplement our analysis by outlining the existing countermeasures and highlighting their limitations. We then propose a new scheme that utilizes a design combining various blocks of the prior solutions and provide more effective defense against selfish mining. We evaluate the workings of our model in  light of our threat model and varying attack conditions. Our proposed scheme is effective in detecting the behavior of a selfish miner and encourages the network to discard his efforts. 

\BfPara{Contributions} In this paper, we make the following contributions. 
\begin{enumerate*}
\item We describe a new form of selfish mining attack by outlining developments of new attack avenues in blockchain community with guaranteed rewards. 
\item We empirically establish the feasibility of this attack by comparing high revenue guarantees with low attack cost. 
\item We outline the limitations of simple and existing countermeasures and propose a new scheme for deterring selfish mining and promoting honest mining practices.
\item We validate our design effectiveness under varying attack conditions and against adaptive attackers. 
\end{enumerate*}

\BfPara{Organization}\label{sec:org} The rest of the paper is organized as follows. In section~\ref{sec:rw}, we review the prior work done to explore selfish mining and its countermeasures. In section~\ref{sec:ps}, we outline the problem statement and preliminaries of our work. In section~\ref{sec:tm}, we describe the threat model and the attack procedure. In section~\ref{sec:cm}, we present our scheme of countering selfish mining, followed by concluding remarks in section~\ref{sec:conclusion}.

\section{Related Work} \label{sec:rw}
Selfish mining in blockchains was identified by Eyal and Sirer \cite{eyal2014majority}, who demonstrated that mining protocols are not incentive-compatible and selfish miners may compromise the system and obtain higher rewards than their due shares. They used a state machine to outline the benefits of selfish mining to malicious miners. As a countermeasure, they proposed a random selection scheme at fork instance to disincentivize the selfish miner. However, in random block selection, the honest miner is equally likely to lose its block during fork and there is no guarantee that the honest miner will win under race conditions. Heliman~\cite{Heilman14}, proposed a ``Freshness Preffered'' (FP) technique in which blocks with recent timestamp are  preferred over old blocks. In FP, when a node is presented with blocks of an honest miner and a selfish miner, it selects the most recent blocks as identified by their timestamps. Although this is an effective technique to spot selfish mining behavior, the information flow in a  peer-to-peer network is not always fair and unexpected delays in the propagation of honest blocks may favor the selfish miner.

\begin{figure}[t]
	\centering
	\includegraphics[width=.8\linewidth]{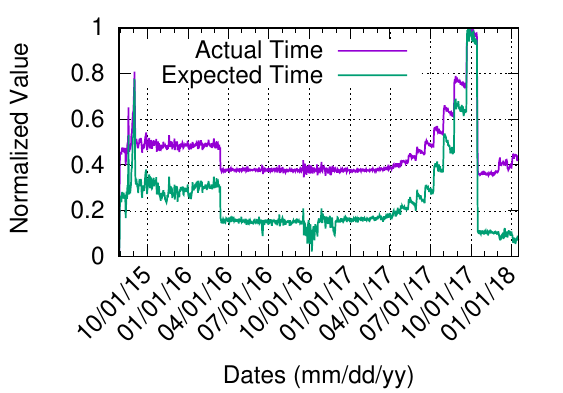}

	\caption{Expected and actual time of blocks published in Ethereum. The non-unifrom delay can be due to varying hash rate or network churn.  }
	\label{et}
\end{figure}

Solat \etal \cite{SolatP16} introduced {\em Zeroblock}, where miners are forced to release their blocks within an expected time. If the miners withhold their blocks for selfish mining and do not broadcast them within the expected time, the peers in the network create their own dummy blocks and append them to their blockchains. However, {\em Zeroblock} is not sustainable in varying hash rate of the network when the difficulty parameter is constant. Expected time of blocks may incur high variance due to hash rate fluctuations which, under {\em Zeroblock}, may invalidate valid blocks. In \autoref{et}, we plot the normalized values of expected block time and the actual block time in Ethereum. Due to varying hash rate or network latency, there is an expected delay in the actual block time which will result in invalidation of valid blocks in {\em Zeroblock}.  Moreover, appending dummy blocks in blockchain creates an additional overhead on the blockchain size, which has reached beyond 163 GB and 450 GB in Bitcoin and Ethereum, respectively. 
Another method to combat selfish mining involves comparing timestamps of transactions in blocks of honest and selfish miners to discover selfish mining behavior. However, miners, as part of the standard practice, prioritize transactions based on fee. Aware of such countermeasures, a selfish miner may include recent transaction of low fee in his block and win the race against an honest miner who mines old transactions with high fee. Also, an adaptive attacker can include fewer or no transactions in his block to avoid the timepstamp checking and still succeed in the attack. Therefore, there is a need for an effective deterrence mechanism to accurately distinguish between adaptive selfish and honest mining.

\section{Problem Statement and Preliminaries} \label{sec:ps}
Over the years, only one selfish mining attack has been reported on a Japanese \cc called Monacoin. Such low prevalence of selfish mining in blockchains can be ascribed to the low probability of success, high attack cost, and low returns. Once a selfish miner finds a competing block in his private chain, he is forced into a race condition in which he competes with the hash rate of the rest of the network to extend his private chain before anyone else appends a block on the main blockchain. As such, the probability of finding a block before the rest of the network becomes a function of selfish miner's hash rate and the aggregate hash rate of the network. To illustrate that, let the probability of success for the selfish miner be  $P(s)$. Let $z$ be the number of blocks that the selfish miner wants to append to his private chain, $\alpha$ be the fraction of selfish miner's hashing power, and $\gamma$ be the fraction of remaining hashing power, where $\alpha + \gamma = 1$. The success probability of selfish miner can then be defined as:
\begin{align}
P(s)= \begin{cases} 
{1} & \text{, $\alpha > \gamma$}\\
\left(\frac{\alpha}{\gamma}\right)^z &\text{, $\alpha < \gamma$}
\end{cases}
\end{align}

\begin{figure*}[t]
\begin{center}
\includegraphics[ width=0.7\textwidth]{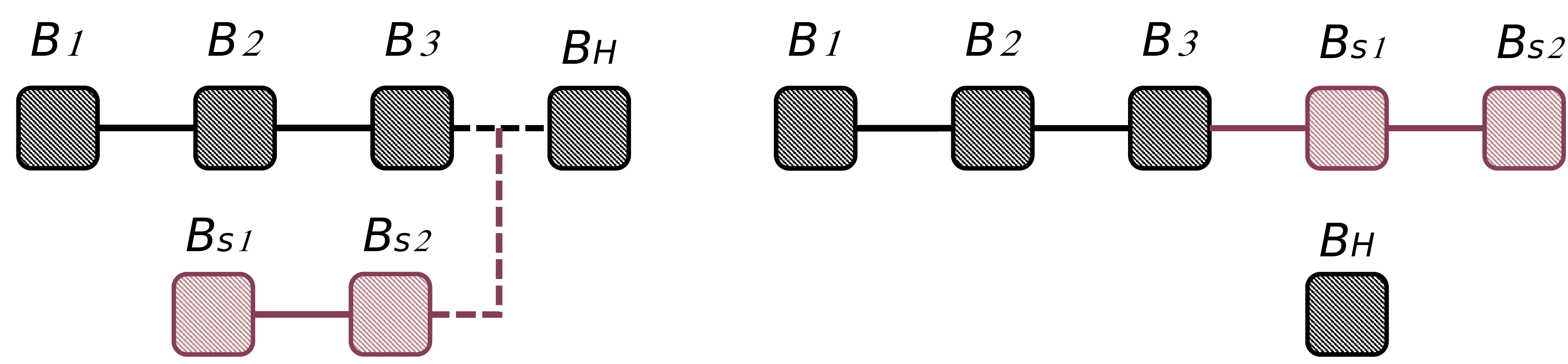}
\caption{An illustration of baseline selfish mining attack in which the selfish miner forks the blockchain. In the start, the honest miner publishes $B_{H}$, which is accepted by the network to elongate the chain. At the same time, the selfish miner computes $B_{S_{1}}$ and $B_{S_{2}}$, and forks the blockchain at $B_{3}$. The parent block of $B_{H}$ and $B_{S_{1}}$ is $B_{3}$. Once forked, the network discards $B_{H}$ and adapts to the longer chain. As a result, the selfish miner succeeds.   }
\label{sm}
\end{center}
\end{figure*}

With low hash rate, a selfish miner may not succeed in launching an attack and is likely to lose the block rewards to the honest miner. To be able to win the race condition with guaranteed returns, the selfish miner needs at least 51\% hash rate to succeed before the rest of the network finds a block. A combination of the 51\% attack and the selfish mining attack will ensure selfish miner's monopoly  over the blockchain. 

However, this strategy has two caveats. First, purchasing hardware to acquire a majority of hash rate in a major \bc is expensive; the mining industry has moved from inexpensive CPU and GPU mining to sophisticated ASIC mining chips that are expensive. Second, as the 51\% hash rate gives the attacker complete control over the network, it can easily be noticed by the network entities and they may discard all blocks published by the selfish miner. Due to these limitations, selfish mining attacks have not been observed frequently in blockchains. 

However, we have noticed recent developments in the \bc market that might facilitate selfish mining in disguise, and prevent peers in the network from discovering such fraudulent activity. Online hashing services, such as NiceHash, have emerged to outsource hashing power to the miners on hourly basis \cite{nicehash_18}. A selfish miner may rent up to 50\% hashing power of a target \bc for a short period of time and carry out successful selfish mining attack. In that case, the attack cost will be the money paid to NiceHash, and rewards will be the block rewards once the private chain is accepted. To calculate the profit of launching such an attack, let $b$ be the block time of a \bc in minutes, $r$ be the block reward for publishing a block, and $c$ be the cost of renting 50\% hash rate of the \bc for one hour. Then, the profit  $p$ for launching a successful attack of $z$ blocks on that \bc can be computed as: 
\begin{equation} \label{equation:increasecost}
  p = \left(z \times r\right) - \left(\frac{z \times b \times c }{60}\right) 
\end{equation}

\begin{figure}[t]
	\centering
	\includegraphics[width=.8\linewidth]{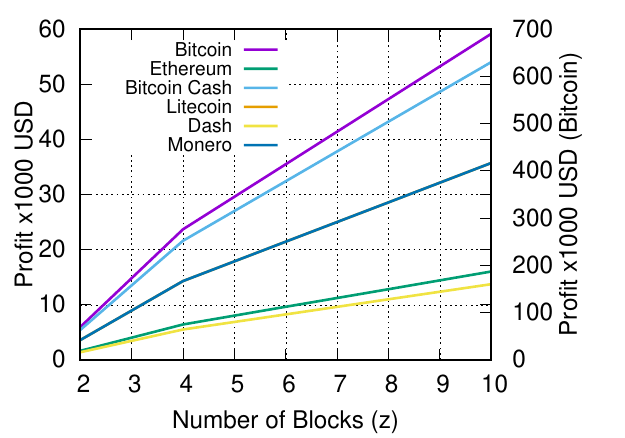}

	\caption{Profit $P$ earned by launching a selfish mining attack of length $z$ blocks on top six cryptocurrencies. Notice that secondary $y$-axis is used for Bitcoin because its Profit margins were high compared to the other cryptocurrencies. }
	\label{Mempool}

\end{figure}

Using \autoref{equation:increasecost}, we calculate the cost to launch a successful attack of two blocks in the top six cryptocurrencies, reported in \autoref{tab:51}. Notice that for each \cc, the block reward is always greater than the attack cost. To launch this attack with a longer private chain, the attacker needs to acquire NiceHash services for a longer duration, which might be more costly and may risk exposing the behavior selfish miner. In \autoref{Mempool}, we show the profit margins of selfish mining for six cryptocurrencies against the length of the private chain of the attacker $z$. Notice that the rewards for Bitcoin and Bitcoin Cash are greater than other cryptocurrencies. Their average block time is 10 minutes, which provides more sustainable attack window compared to Ethereum which has an average block time of 15 seconds. Furthermore, it is interesting to note that the second biggest \cc network, Ethereum, has low rewards and higher attack cost compared to the smaller \cc Litecoin.

\begin{table}[t]
\centering
\caption{Attack cost required and profit margin earned in a selfish mining attack of two block on five major cryptocurrencies. Here, Cap denotes the market cap in USD, cost denotes the attack cost (USD), and Profit denotes the minimum profit earned through block rewards (USD). }
\label{tab:51}
\scalebox{0.90}{
\begin{tabular}{|l|c|l|c|c|}
\hline
{\sc System}    & {\sc CAP} & {\sc Hash Rate} & {\sc Cost} & {\sc Profit} \\ \hline
{\sc Bitcoin}      & 112.7B     & 35,604 PH/s        & 81K  & 69K             \\ \hline
{\sc Ethereum}     & 49.5B      & 222 TH/s           & 1.50K   & 1.6K            \\ \hline
{\sc B.Cash }      & 14.9B           & 5,023 PH/s         & 11.30K   & 5.4K            \\ \hline
{\sc Litecoin}     & 5.7B       & 327 TH/s           & 0.13K    & 3.6K            \\ \hline
{\sc Dash}         & 2.1B       & 2 PH/s             & 0.13K    & 1.4K            \\ \hline
{\sc Monero}       & 2.3B       & 365 MH/s           & 0.10K    & 0.9K           \\ \hline
\end{tabular}}
\end{table}

\section{Threat Model and Attack Procedure} \label{sec:tm}
For our threat model, we assume a selfish miner capable of mining two or more blocks in race conditions. The aim of the selfish miner is to compute valid blocks and withhold them in a private chain to generate a fork against an honest miner. As a result, the attacker would want the network to switch to its longer private chain and discard the block of the honest miner. For that to happen, the attacker would want its private chain to be at least one block longer than the main blockchain to be able to convince the network for a longer proof-of-work and convince them to switch. On the other hand, we assume that the honest miner will follow the conventional mining practices, and will prioritize transactions based on their mining fee. He will try to include as many transactions in the block as possible to gain both block and fee rewards. Furthermore, the honest miner will not withhold his block and  will timely broadcast it to the network upon computation.

\subsection{Baseline Attack} \label{sec:ba}

The baseline attack procedure involves a selfish miner producing two blocks, $B_{S_{1}}$ and $B_{S_{2}}$, and forking the main blockchain to invalidate honest miner's block $B_{H}$. The attacker rents 50\% hash power of Bitcoin from NiceHash for 10 minutes. The attack sequence follows two rounds. In the first round, the attacker computes a first block, $B_{S_{1}}$, using his own hashing power. It then withholds the block and observes the honest miner's block $B_{H}$  being accepted by the network. 

In the second round, the attacker uses the rented hash power to compute the next block $B_{S_{2}}$ before anyone else in the network. Once the block is computed, the attacker forks the main blockchain with its private chain as illustrated in \autoref{sm}. As a result, the network switches to the forked private chain of the selfish miner and discards the block of the honest miner. The selfish miner succeeds in the attack and wins more rewards than the cost incurred in the attack.

{\def\arraystretch{0.85}
\begin{table}[t]\caption{List of notations used in this paper}
\centering
\scalebox{0.85}{
\begin{tabular}{|l c l| } 
\hline
$S_{M}$ & $\defeq$ & Selfish miner\\
$H_{M}$ & $\defeq$ & Honest miner\\
$F_{S}$ & $\defeq$ & State of Fork\\
$N_{S}$ & $\defeq$ & Normal State\\
$Sstate$ & $\defeq$ & Truth state for selfish miner\\
$Fstate$ & $\defeq$ & Future state for selfish miner\\
$Hstate$ & $\defeq$ & Truth state for honest miner\\
$n$ & $\defeq$ & Total blocks computed by selfish miner \\
$B_{S_{i}}$ & $\defeq$ & Selfish miner's blocks, where $i = \{1,2,\ldots,n\} $\\
$X_{S_{i}}$ & $\defeq$ & Height of each block $B_{S_{i}}$, where $i = \{1,2,\ldots,n\} $\\
$B_{H}$ & $\defeq$ & Honest miner's block	\\
$Y_{H}$ & $\defeq$ & Height of $B_{H}$ \\
$p$ & $\defeq$ & Total number of transactions in $B_{S_{i}}$ \\
$Tx_{j}$ & $\defeq$ & Transaction in  $B_{S_{i}}$, where $j = \{1,2,\ldots,p\} $  \\
$E(Tx_{j})$ & $\defeq$ &Expected confirmation height of $Tx_{j}$ \\
$q$ & $\defeq$ & Total number of transactions in $B_{H}$ \\
$Ty_{k}$ & $\defeq$ & Transaction in $B_{H}$, where $k = \{1,2,\ldots,q\}$  \\
$E(Ty_{k})$ & $\defeq$ & Expected confirmation height of $Ty_{k}$ \\
\hline
\end{tabular}}
\label{tab:notations}
\end{table}
}

\section{Countermeasures} \label{sec:cm}
To counter this attack, we introduce the notion of "truth state" for blocks at the fork instant to identify  selfish mining behaviors. We append a parameter of ``expected confirmation height'' in the data structure of a transaction. In blockchains, the height of the block is the index number that denotes its position in the chain. A new block adds the height of the chain by factor of 1. Expected confirmation height is the index number of a future block in which the transaction is likely to be mined, and it depends upon the transaction size, the mining fee, and the size of the memory pool. The mining fee and the transaction size assign a priority factor to the transaction. The priority factor shows the incentive for a miner to select the transaction for his block. If the mining fee is high and the transaction size is small, miners are more inclined to prioritize that transaction for their block. Memory pool in blockchains is a repository that caches unconfirmed transactions. If the memory pool size is large, it creates a transaction backlog and pending transactions have to wait to be mined. 

In Bitcoin, for example, an online service called ``{\em Earn}'' uses Monte Carlo simulation techniques to predict the expected confirmation height of a transaction with 90\% confidence interval \cite{Earn18}. Their simulation parameters take the backlog of transactions, the fee priority of the miners over the last three hours, and the rate of the incoming transactions as inputs. Based on these parameters, {\em Earn} predicts the expected confirmation height and the expected delay for a given transaction. This prediction algorithm can also be applied on the software client of the users so that once  a user generates a transaction, its software client can calculate the expected confirmation height of the block and  append it to the transaction before broadcasting it to the network. Under standard mining, the transaction will likely be mined in the target block with 90\% confidence. Thus, the average expected confirmation height of all transactions in the target block will be equal to the actual block height. This can be used to assign a ``truth state'' to the block and further be used to catch selfish miners who deviate from the standard mining. In the following, we elaborate on our design. We list notations in \autoref{tab:notations}, and provide the description of our design in algorithm 1. 

\begin{algorithm}[t]  \label{algo:design}
\SetAlgoLined\SetArgSty{}
\SetKwInput{KwData}{State}
 \KwData{Fork on blockchain $F_{S}$}
\SetKwInOut{Input}{Inputs}  
\Input{$B_{S_{i}}, B_{H}$;\\
}

$Sstate = X_{S_{n}} - \dfrac{(\sum_{j=1}^{p } {E(Tx_{j})} )}{p}$
     \tcp*[l]{Truth state for selfish miner}
$Fstate = X_{S_{1}} - \dfrac{(\sum_{j=1}^{p } {E(Tx_{j})} )}{p}$
     \tcp*[l]{Future state for selfish miner} \label{alg:fs}
$Hstate = Y_{H} - \dfrac{(\sum_{k=1}^{q} {E(Ty_{k})} )}{q}$
     \tcp*[l]{Truth state for honest miner}

\eIf{($Hstate < Sstate$ \textbf{or} $Fstate$ < 0  )}{
        \textbf{Reject} $B_{S_{i}}$
          \tcp*[r]{Reject selfish miner}
        
}
    { ($Hstate >  Sstate$ ) \tcp*[l]{Circumvention}}  {
            $Asize = 0$   }
        
\ForEach{$p \in B_{S_{i}}$ } {
                $Asize = Asize + p$ \tcp*[l]{Compare number of transactions}
                \eIf { ($q > \dfrac{Asize}{n} $ \textbf{or} $Asize =0$  )} {
                \textbf{Reject} $B_{S_{i}}$ \tcp*[l]{Reject if transactions size is small}}
                {\textbf{Accept} $B_{S_{i}}$\; }}
\SetKwInput{KwData}{State}
 \KwData{Normal State $N_{S}$ }

  \caption{Detecting Selfish Mining Behavior}
\end{algorithm}

\subsection{Selfish Mining Detection}\label{sec:dp}
In the light of our threat model and baseline attack (\autoref{sec:tm}), once the selfish miner publishes his private chain to create a fork, two of his blocks $B_{S_{1}}$ and $B_{S_{2}}$ will have transactions with expected confirmation heights $E(Tx_{j})$. His truth state will be evaluated by subtracting the mean height of all the transactions in the first block $B_{S_{1}}$ from the height of the second block $B_{S_{2}}$. For a selfish miner, the difference in the block height of $B_{S_{2}}$ and the average expected height of transactions in first block $E(Tx_{j})$ will be significant; indicating that the miner withheld the block $B_{S_{1}}$ and did not publish it to the network. The greater the length of the private chain of selfish miner, the higher will be the value of mean expected height of $ X_{S_{n}} - E(Tx_{j})$.  

The truth state of an honest miner's block will be calculated by subtracting the mean height of his transactions $E(Ty_{k})$ from the block height $B_{H}$. Smaller difference in the block height and average expected block height will yield to a greater truth state. This will give advantage to the honest miner as his block will have a higher truth state compared to the selfish miner. In the condition of a fork $F$, all the peers in the network will be required to compute the truth state for the competing
blocks of both miners, and if 51\% peers comply with honesty, the honest miner will win the race condition and selfish miner's private chain will be rejected. The fork state $F$ will be resolved and the network will resume the normal state $N$.

\subsection{Circumventing Detection} \label{sec:cd}
An adaptive attacker may still circumvent detection as follows.   
\begin{enumerate*}
    \item Include transactions with future expected block time in the first block to reduce the difference in the height of the latest block and the mean expected confirmation height of transactions in the first block. 
    \item Include fewer or no transactions in each block to achieve a higher truth state than the honest miner ($Sstate \approx X_{S_{n}}$ and $Sstate < Hstate$). 
\end{enumerate*}

To counter the first technique, we add a future state parameter $F_{state}$ in our algorithm (\autoref{alg:fs}), that verifies if the selfish miner has attempted to include transactions belonging to a future block in the current block. If the selfish miner does that, $Fstate$ value will be less than zero, exposing the nature of transactions in each block. In algorithm 1, if the $Fstate$ value is less than zero, then the private chain is rejected. To counter the second technique, we compare the number of transactions in the blocks of honest and selfish miner. If the average number of transactions in the selfish miner's blocks are less than the honest miner's block, it will expose that the selfish miner has tried to artificially achieve a higher truth state by publishing empty blocks. In our algorithm we reject the private chain if such fraudulent behavior is detected.

\subsection{Exceptional Cases} \label{sec:ec}

Since mining is a lottery-based system, there might be instances where an honest miner finds two blocks within ten minutes and forks the network against another honest miner. In such a situation, the honest miner with the longer chain deserves to win the race condition and should not be accounted for selfish behavior. Our algorithm is flexible for such cases as long as standard mining practices are followed. The honest miner with the longer chain must have sizable transactions in each block and all of them need to have expected confirmation height close to the height of their respective block. If those conditions are met, the honest miner will win the race condition and his private chain will be accepted. Therefore, our algorithm ensures fairness even under such circumstances as long as the standard protocols are followed.

\section{Conclusion and Future Work} \label{sec:conclusion}
In this paper, we introduce a form of selfish mining attack on blockchains, that guarantees high rewards with low cost. We outline the nature of this attack and show its profit margins in top six cryptocurrencies. We survey the prior work and discuss their approach and limitations. 
To counter this attack, we leverage honest mining practices to devise a notion of ``truth state'' for blocks during a selfish mining fork. We assign an expected confirmation height to each transaction to detect selfish mining behavior in the network. Our proposed algorithm effectively deters selfish mining and encourages fair mining practices. In future, we aim to estimate the fee overhead of appending the estimated confirmation height in each transaction as well as the processing overhead of applying our algorithm at the software client. 

\BfPara{Acknowledgement}  This work is supported by Air Force Material Command award FA8750-16-0301. "DISTRIBUTION A.Approved for public release; Distribution unlimited. Case Number 88ABW-2018-5853; Dated 26 Nov 2018"
\bibliographystyle{IEEEtranS}
\bibliography{references,conf}

\end{document}